\documentclass[aps,prl,twocolumn,showpacs,superscriptaddress,amsmath,amssym,floatfix,nofootinbib]{revtex4}
\usepackage{graphicx}
\usepackage{dcolumn}
\usepackage{bm}
\usepackage{color}
\usepackage{xspace}
\usepackage{comment}

\newcommand{\KEPLER}{\textsc{Kepler}}

\begin{document}


\title{Effective Helium Burning Rates and the Production of the Neutrino Nuclei}

\author{Sam M. Austin}
\email{austin@nscl.msu.edu}
\homepage{www.nscl.msu.edu/~austin}
\affiliation{National Superconducting Cyclotron Laboratory,\\
 Michigan State University, 640 South Shaw Lane, East Lansing, MI 48824-1321, USA}
\affiliation{Joint Institute for Nuclear Astrophysics\\University of Notre Dame, Notre Dame, IN 46545, USA}

\author{Christopher~West}
\affiliation{Minnesota Institute for Astronomy,
School of Physics and Astronomy,
The University of Minnesota, Twin Cities,
Minneapolis, MN 55455-0149, USA}
\affiliation{Joint Institute for Nuclear Astrophysics\\University of Notre Dame, Notre Dame, IN 46545, USA}

\author{Alexander~Heger}
\affiliation{Monash Centre for Astrophysics,
School of Mathematical Sciences,
Monash University, VIC 3800, Australia}
\affiliation{Minnesota Institute for Astronomy,
School of Physics and Astronomy,
The University of Minnesota, Twin Cities,
Minneapolis, MN 55455-0149, USA}
\affiliation{Joint Institute for Nuclear Astrophysics\\University of Notre Dame, Notre Dame, IN 46545, USA}

\date{\today}

\begin{abstract}
Effective values for the key helium burning reaction rates,
triple-$\alpha$ and $^{12}$C($\alpha,\gamma$)$^{16}$O, are
obtained by adjusting their strengths so as to obtain the best
match with the solar abundance pattern of isotopes uniquely or
predominately made in core collapse supernovae. These effective
rates are then used to determine the production of the neutrino
isotopes.  The use of effective rates considerably reduces the
uncertainties in the production factors arising from
uncertainties in the helium burning rates, and improves our
ability to use the production of $^{11}$B to constrain the
neutrino emission from supernovae.
\end{abstract}


\maketitle


Uncertainties in the reaction rates for stellar helium burning have
long limited the accuracy with which one can predict nucleosynthesis
in massive stars \cite{wea93,tur07,tur09,tur10,wes13}.  In this letter
we outline a new approach, the use of effective reaction rates (ERR),
obtain a candidate ERR, and apply it to the production of the neutrino
nuclei. It appears that this procedure considerably reduces the
uncertainties in the predictions.

Early work along these lines by Weaver and Woosley \cite{wea93} and by
M.\ M.\ Boyes \cite{boy01} (unpublished, but quoted in \cite{woo07})
concentrated on the reaction rate, $r_{\alpha, \gamma}$, of the
$^{12}$C($\alpha,\gamma$)$^{16}$O reaction.  Boyes used the {\KEPLER}
code \cite{wea78,woo95,rau02,woo02} to calculate the pre-supernova
abundance of nine isotopes ranging from $^{16}$O to $^{40}$C, for
various values of $r_{\alpha, \gamma}$ and found that the smallest
spread in their production factors, as measured by their statistical
variance, $\sigma^2$, occurred for a rate about $1.2$ times that of
Buchmann \cite{buc96}.  This rate was used in most subsequent
calculations with {\KEPLER}.  For details see \cite{woo07}.

Later, Tur, {\it et al.} \cite{tur07} improved this procedure by using
a larger set of stars and taking into account supernova explosive
nucleosynthesis, which modified some of the reference abundances.  The
resulting best value was slightly changed, to $1.3$ times that of
Buchmann \cite{buc96}.  A problem with these approaches was that the
value of the triple-$\alpha$ rate, $r_{3\alpha}$, was fixed at its
experimental value, and since this value was itself uncertain, the
overall validity of the process was difficult to assess.

In these attempts to determine a reaction rate, an implicit
assumption was that uncertainties in the calculations
themselves were substantially smaller than those resulting from
uncertainties in the helium burning reaction rates.  It is,
however, not certain that this is the case, since the
simulations do not include all phenomena that might influence
nucleosynthesis.  Although the {\KEPLER} code can calculate the
effects of rotation and magnetic fields, such calculations are
more cumbersome, and these effects were not included in the
above calculations.  In addition, many reaction rates are
uncertain, as are opacities and mass loss rates.  Perhaps the
most important uncertainties are related to convection ({\KEPLER}
uses the Ledoux criterion), semiconvection, and overshooting.
Helium burning reactions are strong sources of energy in the
star and it well known that a small change in these rates can
have a major influence on nucleosynthesis processes affected by
convection \cite{rau02,tur10}.

These issues are not particular to {\KEPLER} but inherent to
most stellar evolution codes. Imbriani \textsl{et al.}
\cite{imb01} studied the influence on stellar evolution, of
changes in the $^{12}$C({$\alpha$}, {$\gamma$})$^{16}$O
reaction, in combination with variations of the mixing
processes: it appeared that these two uncertainties cannot be
treated separately.  In their work, however, they did not vary
the triple-$\alpha$ rate, and did not follow nucleosynthesis
beyond Zn. Sukhbold \textsl{et al.} \cite{SW14} studied the
sensitivity of stellar structure changes to mixing processes
(semiconvection, overshooting) and compared different stellar
evolution codes; they found that while there are significant
differences in the outcomes using the default values for the
codes, parameters for the mixing physics can be adjusted to
give comparable results.  It seems clear that uncertainties in
the two reaction rates and in the mixing physics are to some
extent intertwined and that all are important.

A possible approach in such a situation is to view the
operators as ``effective'', with their parameters fixed by
comparing the results of calculations to data.  One example of
this approach is the use of effective interactions in the
description of nuclear structure using the nuclear shell model
\cite{bro06}. The effective interaction is determined by fitting
low lying energy levels of a set of nuclei.
This procedure has been remarkably successful, and is the basis
of most modern large basis shell model calculations. In many
cases one cannot show in detail why the process works well; its
justification lies in the fact that the procedure works for
many observables.

In this letter we describe a first attempt to obtain effective
reaction rates (ERR) for the helium burning reactions and to apply
them to the production of the neutrino isotopes $^7$Li, $^{11}$B,
$^{19}$F, $^{138}$La, and $^{180}$Ta.  We obtain the ERR by fitting
the production of intermediate-mass and \textsl{s}-only isotopes,
taking advantage of the extensive supernova calculations of West {\it
  et al.}  \cite{wes13}.  In that work, {\KEPLER} was used to model
the evolution of a group of 12 stars (initial masses $12$, $13$, $14$,
$15$, $16$, $17$, $18$, $20$, $22$, $25$, $27$, and
$30\,\mathrm{M}_{\odot}$) from central hydrogen burning to core
collapse; a piston placed at the base of the oxygen shell was
then used to simulate the explosion yielding a total kinetic energy of
the ejecta of $1.2\,\times 10^{51}\,$erg.  The calculations were
carried out for a matrix of rates, covering $\pm {2\sigma}$ for both
$r_{\alpha, \gamma}$ and $r_{3\alpha}$ (176 rate pairs).  This
involved a total of $12 \times 176 = 2112$ simulations.  In the
following discussion the rates are characterized by a multiple of the
standard values as is described in Tur, {\it et al.} \cite{tur07}.
The results were then averaged over a Salpeter initial mass function
(IMF). For each reaction pair, the \textsl{standard deviations} of the
IMF averaged production factors of the intermediate-mass isotopes
($^{16}$O, $^{18}$O, $^{20}$Ne, $^{23}$Na, $^{24}$Mg, $^{27}$Al,
$^{28}$Si, $^{32}$S, $^{36}$Ar, $^{40}$Ca) and the \textsl{s}-only
isotopes ($^{70}$Ge, $^{76}$Se, $^{80}$Kr, $^{82}$Kr, $^{86}$Sr,
$^{87}$Sr) were obtained, and the standard deviations from each of the
two isotope lists were averaged, thereby giving equal weight to the intermediate-mass and s-only isotopes.

\begin{figure}
\centering
\includegraphics[angle=90,width=\columnwidth]{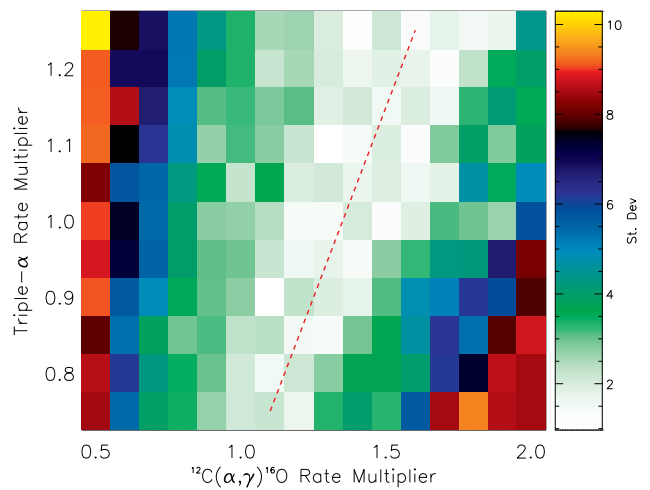}
\caption{The standard deviations for the IMF average of the production
  factors of the intermediate mass and weak \textsl{s}-only isotopes, including
  explosive yields and yields from stellar winds. Only models that satisfied the compactness condition $\xi_{2.5} < 0.25$ were
  included.  The dashed line ($r_{\alpha, \gamma} = r_{3\alpha} + 0.35$) was drawn through the overall minima of
  the calculated standard deviations. Adapted from Fig.~13 of
  \cite{wes13}\label{fig:w13}}
\end{figure}

Before a comparison to observed abundances two corrections were
made. First, models that were likely to collapse to a black
hole were filtered out by including only models with a
compactness factor \cite{oco11, woo12, SW14} satisfying
$\xi_{2.5} < 0.25$.  In addition, the observed \textsl{s}-only
abundances were corrected for the contributions of other
processes.  For details of these calculations see West {\it et
al.} \cite{wes13}. The results are shown graphically in
Fig.~\ref{fig:w13}.

These results are somewhat surprising.  We expected that both
the rates would be individually constrained, but instead we
find that the best fit points lie within a band.  A large range
of $r_{3\alpha}$ is allowed, but the relationship between
$r_{\alpha, \gamma}$ and $r_{3\alpha}$ is constrained.  A best
fit curve is shown. It passes through the overall minimum of
points, each of which is the minimum local standard deviation
of the fitted abundances of intermediate mass and weak
\textsl{s}-isotopes determined as described above.  There is no
strong reason for choosing one point on the ERR rate line over
another; this line is taken as the best available description
of the ERR.  Clearly, the best value of one of the rates
depends on what the other rate is chosen to be. If a new
measurement showed reliably that the actual value of
$r_{3\alpha}$ was $1.2$ ($0.8$) instead of $1.0$, one would
choose a significantly larger (smaller) value of $r_{\alpha,
\gamma}$ to best predict the nucleosynthesis of the
intermediate-mass and weak-\textsl{s} isotopes; the range of
values is $35\,\%$.

We now apply this ERR to study the production of $^7$Li, $^{11}$B,
$^{19}$F, $^{138}$La, and $^{180}$Ta in the neutrino process.  The
fundamental picture is simple: neutrinos emitted by the proto-neutron
star resulting from core collapse interact with relatively abundant
nuclei in the stellar envelope to form the precursors of the neutrino
isotopes.  After decay and processing in the ensuing shock wave, these
become the observed isotopes.  Austin, {\it et al.}  \cite{aus11}
concluded that production of $^{11}$B in the neutrino process might
serve to constrain the average neutrino production in supernovae.  The
uncertainties arising from the uncertainties in the helium burning
rates, however, were relatively large, and it seemed unrewarding to
pursue the issue until one had a better handle on the helium burning
rates.

\begin{figure}
\center
\includegraphics[width=\columnwidth,clip]{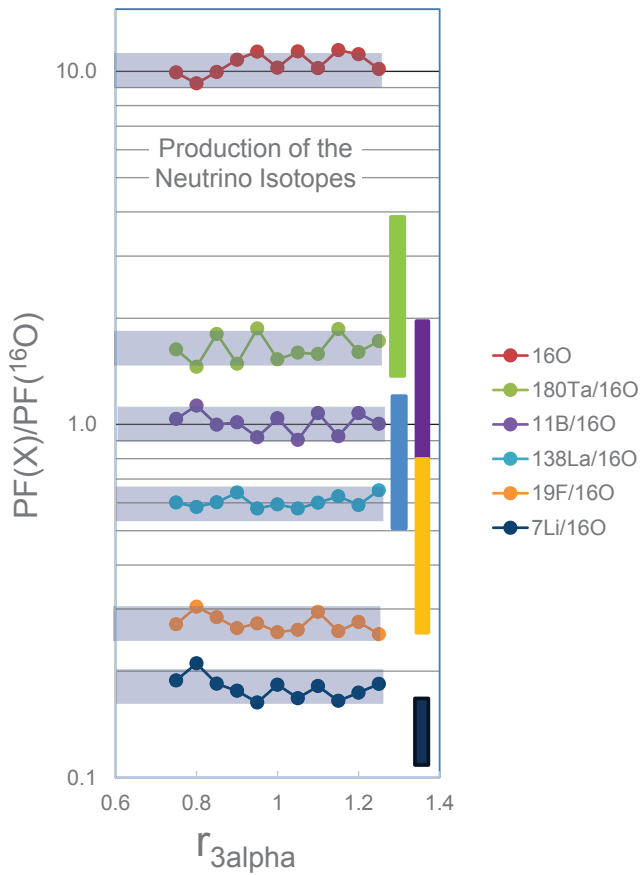}
\caption{Production factors for the neutrino isotopes normalized to
  those for $^{16}$O. The abscissa is the position along the ERR line
  of Fig.~\ref{fig:w13}, parameterized by the value of
  $r_{3\alpha}$. The narrow bands covering $\pm 10\%$ give an
  indication of the precision of the present results.  The bars on the
  right of the graph show the spread of uncorrelated errors obtained
  in Austin, {\it et al.} \cite{aus11}.  As noted in that paper, the
  values of the production ratio for $^{11}$B that would agree with
  observation is about $0.4$.\label{fig:pf}}
\end{figure}

We followed the general procedures outlined in Heger, {\it et
al.}, and Austin, {\it et al.}, \cite{heg05,aus11}, but
calculated the production of the neutrino isotopes for ten ERR
points along the best fit curve of Fig.~\ref{fig:w13}. We used
Fermi-Dirac neutrino spectra, with temperatures for
$\nu_\mathrm{e}$, $\bar{\nu}_\mathrm{e}$, and $\nu_\mathrm{x}$
of $4\,$MeV, $5\,$MeV and $6\,$MeV; $x$ stands for $\mu$ and
$\tau$. The results are shown in Fig.~\ref{fig:pf}. For the
entire range of the ERR line of Fig.~\ref{fig:w13}, the
deviations from a constant value are typically $\pm 10\%$ or
less.  The remaining variations arise (at least mainly) from
binning and aliasing effects, because the ERR line does not
pass precisely through the values of $r_{\alpha, \gamma}$ and
$r_{3\alpha}$ used in the models. Note that in Fig. 2, as well as in Figs. 3 and 4, the values of $r_{3\alpha}$ shown on the abscissa lie along the ERR line, and hence describe implicitly the values of $r_{\alpha, \gamma}$.

This is to be compared to the much larger ranges found in Austin, {\it
  et al.} \cite{aus11} when uncorrelated uncertainties of $r_{\alpha,
  \gamma}$ and $r_{3\alpha}$ were considered.  These uncertainties are
also shown as bars near the right-hand ordinate of
Fig.~\ref{fig:pf}. It is not straightforward to assess the
accuracy of the ERR.  We obtained a rough estimate of possible
effects of moving the line to the left by changing $r_{\alpha,
\gamma}$ by $-0.1$, corresponding to the ERR that is obtained by fitting the intermediate isotopes only. This changed the production of the
neutrino isotopes by between $5\,\%$ and $12\,\%$.

One must ask whether these encouraging results are reliable.  As a
minimum, the use of ERRs allows one to deal with the effects of
uncertainties in the reaction rates and the weaknesses of the model
calculations in a unified way.  It is striking that the production
factors for the neutrino nuclei, which owe their origins to different
shells in the star \cite{woo90,heg05}, vary so little with position
along the line.

\begin{figure}
\center
\includegraphics[clip=True,width=\columnwidth]{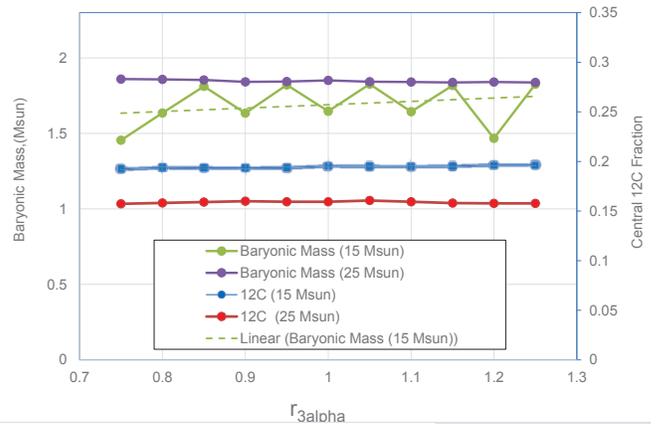}
\caption{Values of the central carbon fraction at the end of helium
burning (right-hand ordinate)and of the baryonic mass of the progenitor
of the stellar remnant (left-hand ordinate)\cite{wes13}.
\label{fig:cmass}}
\end{figure}

Another striking qualitative feature, shown in
Fig.~\ref{fig:cmass}, is that the values of the central mass
fraction at the end of helium burning are nearly constant along
the ERR line.  A similar statement, see Fig.~\ref{fig:cmass},
can be made for the baryonic mass of the progenitor of the
remnant of that star.  The larger variability for the
$15\,\mathrm{M}_{\odot}$ star apparently reflects a sensitivity
to small changes in the reaction rates \cite{SW14} that cannot
be described by an ERR.  There is, however, only a very weak
overall trend with the value of $r_{3\alpha}$.  It has been
pointed out \cite{imb01} that advanced stages of evolution are
strongly influenced by the central $^{12}$C abundance at the
end of helium burning.  The constant value of the carbon
fraction along the ERR line may then provide an understanding
of why the ERR apparently works well.  See also the detailed
study of Sukhbold \textsl{et al.} \cite{SW14}.

\begin{figure}
\center
\includegraphics[clip=True,width=\columnwidth]{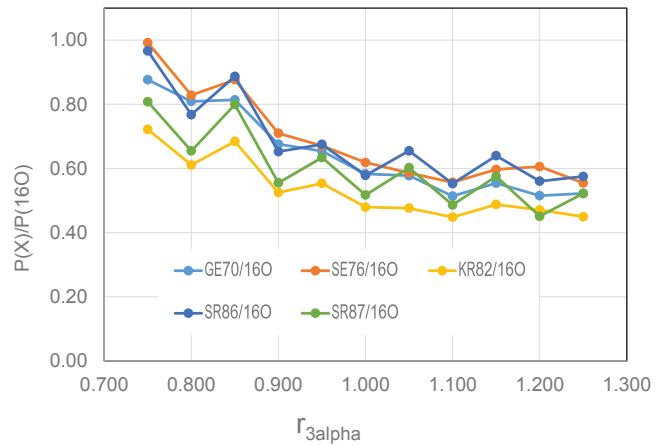}
\caption{The production factors of the \textsl{s}-only nuclei.
 $^{80}$Kr is omitted because {\KEPLER} does not treat the $T$
 dependence of $^{79}$Se decay which affects $^{80}$Kr production.
\label{fig:sonly-pf}}
\vskip -\baselineskip
\end{figure}

\begin{figure}
\center
\includegraphics[angle=90,width=\columnwidth]{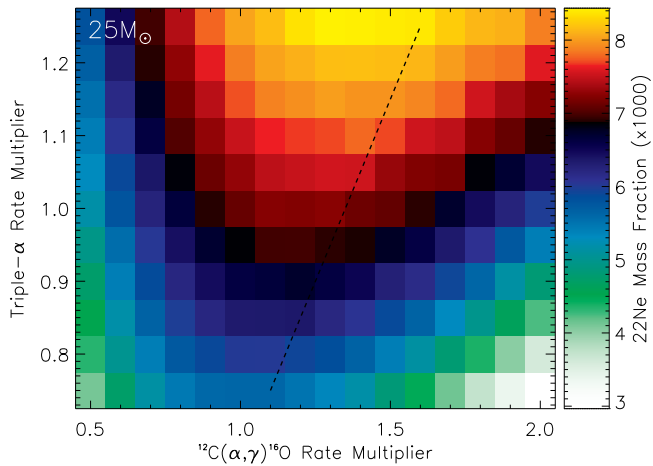}
\caption{Amount of $^{22}$Ne left in the core after end of central He
burning for $25\,\mathrm{M}_{\odot}$ stars. The dashed
  line indicates the ERR valley from Figure~\ref{fig:w13}.
\label{fig:22ne}}
\end{figure}

It may be, however, that fitting additional information could
provide a better ERR, or illuminate other processes. With this
in mind we examined the production factors of the intermediate
and \textsl{s}-only isotopes. The intermediate isotopes provide
no obvious additional information, but the \textsl{s}-only
isotopes do. We find (Fig.~\ref{fig:sonly-pf}) that for larger
$r_{3\alpha}$, their average production factor, normalized to
that of $^{16}$O, is smaller, decreasing significantly from
$\textrm{0.85}$ and reaching a plateau of $\textrm{0.55}$ for
$r_{3\alpha}$ $\gtrsim$ $\textrm{1.0}$.  The standard
deviations of the production factors, fitted in deriving the
ERR, do not change significantly.  This behavior arises from
the temperature sensitivity of the
$^{22}$Ne$(\alpha, n)^{25}$Mg reaction.  At lower
$r_{3\alpha}$, the burning temperature at the end of helium
burning is higher \cite{tur09}, and so is the
$^{22}$Ne$(\alpha, n)^{25}$Mg reaction rate, resulting in
more destruction of $^{22}$Ne as shown in Fig.~\ref{fig:22ne}, more neutron production, and a stronger weak \textsl{s}-process. We note, however, that the $^{22}$Ne burning rates are quite uncertain, so the effect may be stronger or weaker than shown.

Assuming that we have a sufficient understanding of the
observed \textsl{s}-only abundances, and of the weak
s-process, these observation favor helium burning rates at
the lower end of the present ERR. Caution is warranted, however, because of the need to correct for the significant contribution of other
processes to the \textsl{s}-only isotopes,the small number
of these isotopes \cite{wes13b}, and the uncertainty in the $^{22}$Ne$(\alpha, n)^{25}$Mg rate.

To summarize, we have made a first attempt at developing an ERR
for the two helium burning reactions, based on minimizing the
standard deviation in the production factors of two groups of
isotopes: the intermediate mass isotopes and the
\textsl{s}-only isotopes.  This results in a correlation
between the best values of the
$^{12}$C($\alpha,\gamma$)$^{16}$O and triple-$\alpha$ rates. We
have taken this representation of the ERR, as shown in
Fig.~\ref{fig:w13}, and evaluated the production of the
neutrino nuclei at various points along the ERR line. They are
essentially the same at all ERR points, lending credence to the
procedure used to determine the effective rates. The success of
the ERR may be related to the fact that the central $^{12}$C
densities and remnant masses along the ERR line (for a
$15\,\mathrm{M}_{\odot}$ and $25\,\mathrm{M}_{\odot}$ stars)
are very closely the same.

These results apparently remove what was a major hurtle to
comparing neutrino isotope abundances to the predictions: that
they depended so strongly on poorly known helium burning rates.
It now becomes meaningful to address other uncertainties: the
explosion energy, the neutrino interaction cross section, the
cross sections for reactions that process the mass-11 products,
and the nature of the neutrino spectrum, as outlined in Austin
{\it et al.}  \cite{aus11} to see whether, as described there,
one can use the abundance of $^{11}$B to determine the average
emission of neutrinos in supernova explosions.

We note that the derivation of the ERR depends on the model and
is only valid for {\KEPLER} and the specific values used for
input physics, including mixing processes, reaction rates,
initial abundances and metallicity.

We thank Stan Woosley for advice and discussions about this
manuscript.  Research support is from: US NSF; grants PHY08-22648
(JINA), PHY11-02511; and US DOE: contract DE-AC52-06NA25396,
grants DE-FC02-01ER41176, FC02-09ER41618 (SciDAC),
DE-FG02-87ER40328. AH was supported by an ARC Future Fellowship
(FT120100363).



\end{document}